# Theoretical prediction of the source-detector separation distance suited to the application of the spatially resolved spectroscopy from the near-infrared attenuation data cube of tissues

Yong-Wu Ri[1], Sung-Ho Jong[1], and Song-Jin Im[2]

[1]Department of Chemistry and [2]Department of Physics,
**Kim Il Sung** University, Daesong district,
Pyongyang, DPR Korea

**Abstract**: The modified Beer-Lambert law (MBL) and the spatially resolved spectroscopy are used to measure the tissue oxidation in muscles and brains by the continuous wave near-infrared spectroscopy. The spatially resolved spectroscopy predicts the change in the concentration of the absorber by measuring the slope of $A(\rho)$ according to the separation $\rho$ and calculating the absorption coefficients of tissue on the basis of the slop in attenuation at the separation distance satisfying the linearity of this slop. This study analyzed the appropriate source-detector separation distance by using the diffuse approximation resolution for photon migration when predicting the absorption coefficient by the spatially resolved spectroscopy on the basis of the reflective image of the tissue. We imagine the 3 dimensional attenuation image with the absorption coefficient, reduced scattering coefficient and separation distance as its axes and obtained the attenuation data cube by calculating the attenuation on a certain interval of coordinate on the basis of the diffuse approximation expression. We predicted the separation distance appropriate for the application of the spatially resolved spectroscopy by calculating and analyzing the first derivatives and second derivatives of attenuation with respect to the coordinates and also doing the differential pathlength factors and first derivatives of the attenuation with respect to the absorption coefficient from the attenuation data cube. When analyzing the hemoglobin derivatives in tissues, the appropriate separation distances are 3-5cm and the value of its corresponding differential pathlength factors are from 3.5 to 5. These data agree with the preceding experimental data.

Key words:
Near-infrared spatially resolved spectroscopy, source-detector separation distance, differential pathlength factor, attenuation data cube

## 1. Introduction

About 20 years have passed since the near-infrared spectroscopy was applied to determine the amount of the hemoglobin derivatives and this technique has developed to the high level enough to judge even the functional activation of the brain cortex of human being[1]. Although this technique is wide in use and diverse in application, the simplest and most general physical model which is the basis of this technique is the model of the single detector whose sample is dealt with the semi-infinite medium. In such case the light source and detector are placed in the air against the surface of tissue (figure 1).

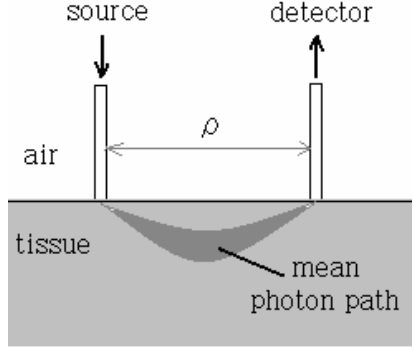

Figure1. Model of the semi infinite medium

The diffusion theory of the continuous light for the semi infinite media predicts the diffusion reflectivity $R(\rho; \mu_a, \mu_s')$ of the stationary state as a function of ρ, where ρ is the source-detector separation along the sample surface[2].

In such case, $R(\rho; \mu_a, \mu_s')$ is the flux of the diffuse photons flux coming from the boundary interface of the tissue.

$$R(\rho, z_0) = \frac{1}{4\pi}\left(z_0(\mu_{eff} + \frac{1}{r_1})\frac{\exp(-\mu_{eff} r_1)}{r_1^2} + (z_0 + 4BD)(\mu_{eff} + \frac{1}{r_2})\frac{\exp(-\mu_{eff} r_2)}{r_2^2}\right) \quad (2)$$

where $r_1 = (z_0^2 + \rho^2)^{1/2}$, $r_2 = ((z_0 + 4BD)^2 + \rho^2)^{1/2}$, $D = 1/(3(\mu_a + \mu_s'))$,

$$\mu_{eff} = (\mu_a/D)^{1/2}, \quad z_0 = (\mu_a + \mu_s')^{-1}.$$

And B is the constant related with the total internal reflection and depends on the refractive index of the medium. Equation 2 agrees with the result of the spatially resolved Monte Carlo simulations in case the separation distance is long. The improved solution was proposed in order to characterize more correctly the optical properties of tissue[3].

$$R_{imp} = 0.118\phi(\rho, z_0) + 0.306 R(\rho, z_0) \quad (3)$$

$$\phi(\rho, z_0) = \frac{1}{4\pi D}\left[\frac{\exp(-\mu_{eff} r_1)}{r_1} - \frac{\exp(-\mu_{eff} r_2)}{r_2}\right] \quad (4)$$

The Equation (2)-(4) provides the direct analysis method to calculate the diffuse reflection at the separation distance given the concentration of absorbers and the coefficient of the reduced scattering. The main absorbers in the tissue are oxy- and deoxyhemoglobin and water.
The reduced scattering coefficient depends on the concentration of the cell and the refraction index of the surrounding media. When the medium is homogenous, it is possible to calculate the oxygen concentration of the tissue by measuring the absorption and scattering coefficients as a function of light wavelength and by solving the following simultaneous equation.

$$\mu_a(\lambda) = \sum_i \varepsilon_i(\lambda) \cdot C_i$$

Where, $\varepsilon_i(\lambda), C_i$ are the molar absorption coefficient and concentration respectively.

In given wavelength and in separation distance ρ, light attenuation A(ρ) is defined as follows:
$$A(\rho) = -\log_{10} R(\rho) \quad (5)$$

Using the modified Beer-Lambert law (MBL)[4], the above expression can be rewritten as follows:
.
$$A = \beta \cdot \mu_a + G \quad (6)$$

$$\beta = \frac{\partial A}{\partial \mu_a} = L = DPF \cdot \rho \quad (7)$$

Where, L is the photons average pathlength of the photons (effective optical pathlength), G is light attenuation due to only scattering in the tissue, and DPF is the dimensionless differential pathlength factor, a multiplier to account for the optical pathlength increased due to scattering. It is impossible to obtain the absolute concentration of the chromophores in the medium by using the expression (6) due to the scattering factor G. But if G is constant in the given experimental condition and we know DPF, that is, β, it is possible to obtain the change in the concentration of the tissue chromophores on the basis of the change in the attenuation. If we don't know β, it is possible to obtain the change in the relative concentration on the basis of $\beta \cdot \Delta c$

$$\Delta A = \beta \cdot \Delta \mu_a \quad (8)$$

$$\Delta c = \left(\frac{1}{\beta}\right) \cdot \varepsilon^{-1} \cdot \Delta A \quad (9)$$

Where, ε is a molar absorption coefficient of the absorption components.

Now MBL and the spatially resolved spectroscopy[5,6] are used to measure the degree of oxidation of the tissue in the skin and brain by the continuous wave near infrared spectroscopy. By the spatially resolved spectroscopy we measure the slop of A(ρ) according to the separation distance ρ and calculate the absorption coefficient $\mu_a$ in the following expression when approximating the reduced scattering coefficient as the function of the wavelength.

$$\frac{\partial A}{\partial \rho} = \frac{1}{\ln 10}((3\mu_a \cdot \mu_s')^{1/2} + \frac{2}{\rho}) \quad (10)$$

But this method can be applied effectively only when the separation distance is much greater than the reciprocal of the reduced scattering coefficient and $\partial A / \partial \rho$ is constant with ρ.

In most detectors used in determining the amount of oxygen in the tissue, the separation distance is constant. This has certain limitation on increasing the range of measure. Because it is necessary to set the separation distance suited to the absorption coefficient and reduced scattering coefficient of the sample to enhance the precision of the measurement. In the meantime, CCD camera is used as a detector to study the hemodynamic change in the tissue by the near infrared spectroscopy or to obtain the penetration and reflection images of the tissue[7,8]. In such case, since each pixel of the camera acts as a detector, the source-detector separation distances vary from all the pixels. What is important in enhancing the precision of the measurement is to find the proper pixel for the analysis.

We calculated the attenuation data cube with the absorption coefficient, reduced scattering coefficient of tissue and separation distance of detector as its independent variables using the diffuse approximation resolution for the photon migration. And by using the attenuation data cube, it predicted the source-detector separation distances necessary to determine the absorption coefficients from the reflection image of the tissue by the spatially resolved spectroscopy.

## 2. Calculation method

**The attenuation (or the reflection strength) in the CCD reflection image of the tissue obtained from the single light source is a function of the distance from the centre of the light source to the pixel.**

In case of measuring such image with the different samples and wavelengths, it is possible to obtain the multi layer reflection image..

When dealing with such multilayer reflection image, we should imagine the hypercube. This hypercube A(ρ, $\mu_a$, $\mu_s'$) consists of the 3 space directions(absorption coefficient $\mu_a$, reduced scattering coefficient $\mu_s'$, source-detector separation distance ρ(these correspond to the voxel coordinates) and the attenuation direction A. We represented this hypercube again as the 3 dimensional data cube. In data cube each voxel value represents attenuation measured by single detector for given wavelength and given tissue.

If expressing the light attenuation as the data cube, it is possible to analyze the attenuation while considering both the kind of sample and type of detector and also to analyze the reflection image according to any separation distance and the spectrum image according to the wavelength. And it is possible to predict the separation distance suited to the application of the spatially resolved spectroscopy.

To calculate the attenuation data cube, first of all, we set the value interval of the absorption coefficient $\mu_a$ as $0.01$-$0.5 \text{cm}^{-1}$, the value interval of the reduced scattering $\mu_s'$ as $5$-$15 \text{cm}^{-1}$ that the various tissues can have and the value interval of ρ as $0.1$-$5\text{cm}$.

We calculated the light attenuation data cube $A(\rho, \mu_a, \mu_s') = -\log_{10}(R(\rho, \mu_a, \mu_s'))$ using the expression (2). And then we calculated the data cube $\partial A/\partial \rho, \partial A/\partial \mu_a, \partial A/\partial \mu_s'$ showing the slopes of attenuation with respect to the directions of the voxel coordinates ρ, $\mu_a, \mu_s'$, respectively. And we calculated the data cube of DPF using the following relationship

$$\frac{\partial A(\rho, \mu_a, \mu_s')}{\partial \mu_a} = L(\rho, \mu_a, \mu_s') = DPF \cdot \rho$$

To confirm the voxels satisfying the linearity of the attenuation slope with respect to the separation distance $\partial A/\partial \rho$, we calculated the second derivatives $\frac{\partial}{\partial \rho}\left(\frac{\partial A(\rho, \mu_a, \mu_s')}{\partial \rho}\right)$ of the attenuation $A(\rho, \mu_a, \mu_s')$ with respect to the separation distance and determined the distribution of the voxel values approaching zero in the data cube.

All the calculations and drawings were made using the MATLAB functions.

### 3. Results and consideration
#### 1) Space distribution of the attenuation values

The space distribution of the attenuation values in the attenuation data cube $A(\rho, \mu_a, \mu_s')$ and the distribution of pixels according to the attenuation values are shown in figure 2 and 3. In the figure 2, left figure shows the distribution of the pixel values in the plane rotated on the axis of the separation distance at a certain angle and the right figure shows the distribution in the boundary planes and planes perpendicular to the axis of the reduced scattering coefficient.

As shown in the figure, the longer separation distance and the greater absorption coefficient and reduction scattering coefficient are, the greater attenuation value is (the less reflection strength is) and in case the separation distance is long, the greater absorption coefficient is, the greater attenuation value is. And data positions (voxels) with the attenuation values between 2-8 are distributed mainly in the data space.

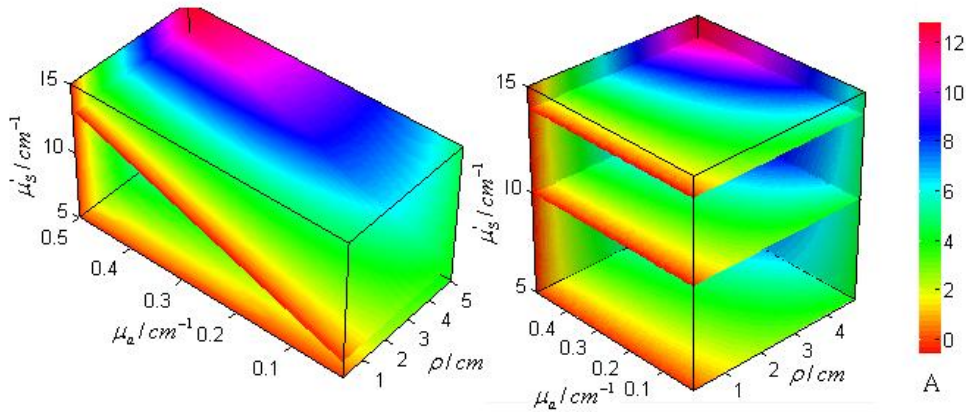

Figure 2. Space distribution of the attenuation values.

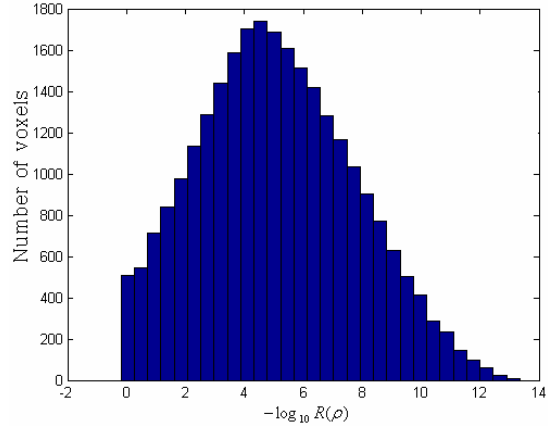

Figure 3. Distribution of the voxels according to the attenuation values.

The negative values appear among the calculated attenuation values in case the separation distance is very short because the approximate resolution (eq. 2) of the diffusion equation can be applied only when the separation distance is much longer than the free pass length of the photon and the absorption coefficient is much less than the reduced scattering one[6,9].

Since the absorption and reduced scattering coefficients change according to the wavelength and the concentration of absorber, all the measured attenuation values are contained in this data cube, regardless of the kinds of the tissue and the concentration of the absorbers. Since the approximate resolution (eq. 2) of the diffusion equation is true only in case the separation distance is more than 1, the attenuation values measured by most detectors are ones (2-8) of the range showing the green color in figure..

**2) Space distribution of the attenuation slope values**

The distribution of the attenuation slope values according to the pixel coordinate $\rho, \mu_a, \mu_s'$ is shown in figure 4. These values help to assess the effect of the change in the direction of the coordinate on the change in the attenuation at any pixel (one condition for the measurement) and to judge the range where that effect gets constant. It is shown that the change in the attenuation slope $\partial A/\partial \rho$ value according to the separation distance gets greater in case the absorption coefficient is little and the separation distance is short but it gets slow as the separation distance gets longer. And it is shown that the effect of the change in the absorption coefficient (change in the concentration of absorber or one of the wavelength) on the attenuation is the greatest and the effect of the change in the reduced scattering coefficient on the attenuation is very little.

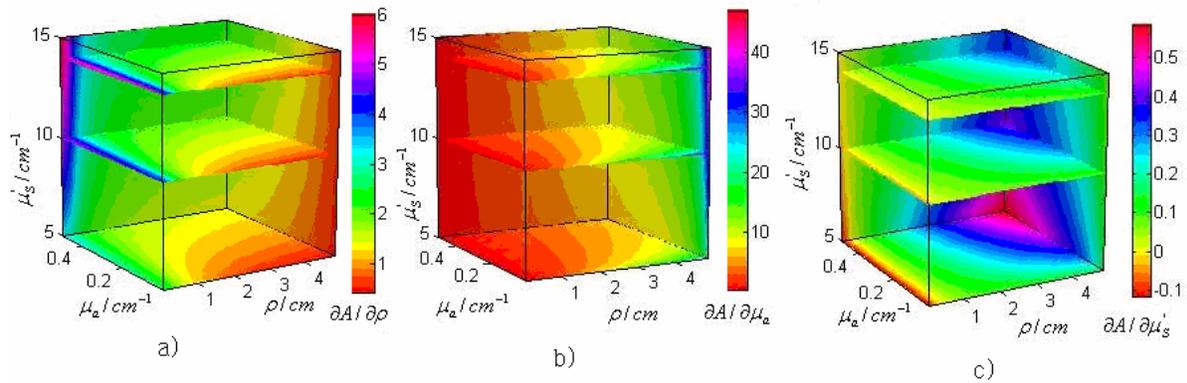

Figure4. Space distribution of the attenuation slope values;
a) $\partial A/\partial \rho$ ,b) $\partial A/\partial \mu_a$ ,c) $\partial A/\partial \mu_s'$

### 3) Prediction of the separation distance suited to the application of the spatially resolved spectroscopy

To determine the absorption coefficient using the expression (10), the attenuation slope values of the separation distance must be constant in a certain separation distance interval. Figure 5 shows the result of the calculation of the second derivative data cube $\partial^2 A/\partial \rho^2$ of the attenuation.

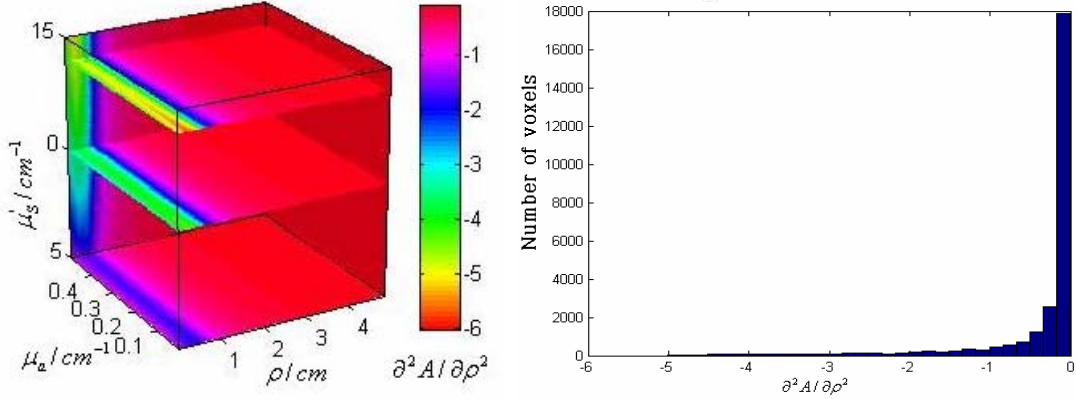

Figure 5.   Space distribution of the second derivative $\partial^2 A/\partial \rho^2$ values of the attenuation

Table 1. Average values of $\partial^2 A/\partial \rho^2$ in a certain separation distance interval

| ρ/cm | $\mu_a$=0.01～0.5cm$^{-1}$ $\mu_s^{'}$=5～15 cm$^{-1}$ | $\mu_a$=0.01～0.2cm$^{-1}$ $\mu_s^{'}$=10～15 cm$^{-1}$ | $\mu_a$=0.01～0.1cm$^{-1}$ $\mu_s^{'}$=10～15 cm$^{-1}$ |
|---|---|---|---|
| 1～2 | -0.340 | -0.371 | -0.386 |
| 2～3 | -0.117 | -0.123 | -0.128 |
| 3～4 | -0.058 | -0.061 | -0.062 |
| 4～5 | -0.035 | -0.036 | -0.037 |

As shown in figure 5, at most voxels of the data cube the values of the second derivative are between -1 and 0. To find the more proper separation distance, we calculated the average value of $\partial^2 A/\partial \rho^2$ in a certain separation distance interval (Table 1). As shown in table1, when the separation distance is greater than 3, the values of the second derivative of the attenuation are distributed around zero. This shows that when applying the spatially resolved spectroscopy using the diffuse approximation resolution for photon migration , it is possible to enhance the approximation degree only when changing the separation distance according to the sample condition (absorption coefficient and reduced scattering coefficient).

### 4) Space distribution of DPF values

The distribution of DPF values according to the pixel coordinate is shown in figure 6.

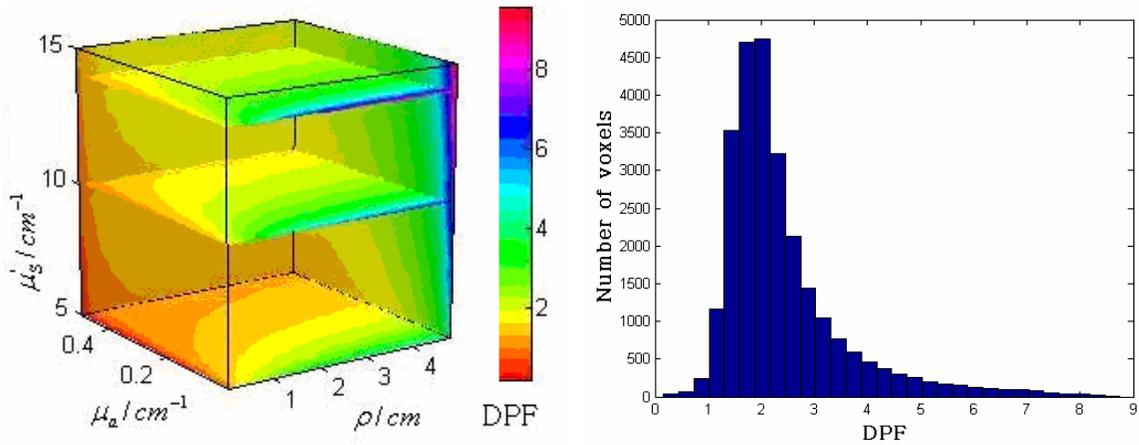
Figure 6. Space distribution of DPF values and one of the voxels

From figure 6, we can see the differential pathlength factor not only depends on the separation distribution and the reduced scattering coefficient but also the absorption coefficient of the medium to a certain degree. In principle, the differential pathlength factor values must be greater than one but there are values less than one in the calculated result. These values appear in case the separation distance is near zero and the absorption coefficient is great. This means that if the diffuse approximation resolution used for the calculation is shorter than the separation distance, that resolution is not true.

The table 2 shows the average values of the differential pathlength factor at the separation distance interval where the spatially resolved spectroscopy can be applied.

Table2. Average values of DPF at a certain separation distance interval

| ρ/cm | $\mu_a$=0.01~0.5cm$^{-1}$ $\mu_s'$=5~15 cm$^{-1}$ | $\mu_a$=0.01~0.2cm$^{-1}$ $\mu_s'$=10~15 cm$^{-1}$ | $\mu_a$=0.01~0.1cm$^{-1}$ $\mu_s'$=10~15 cm$^{-1}$ |
|---|---|---|---|
| 1~2 | 2.27 | 3.45 | 4.17 |
| 2~3 | 2.51 | 3.93 | 4.86 |
| 3~4 | 2.64 | 4.18 | 5.24 |
| 4~5 | 2.72 | 4.33 | 5.48 |

As shown in figure 2, in case the separation distance is greater than 2, DPF has value of 3~5.

**5) Explanation of the analysis condition of the hemoglobin derivatives in the muscle tissue**

Figure 7 shows the measured result of the extinction coefficient of the oxyhemoglobin($O_2$Hb) and deoxyhemoglobin(HHb) in the tissues[10]. .

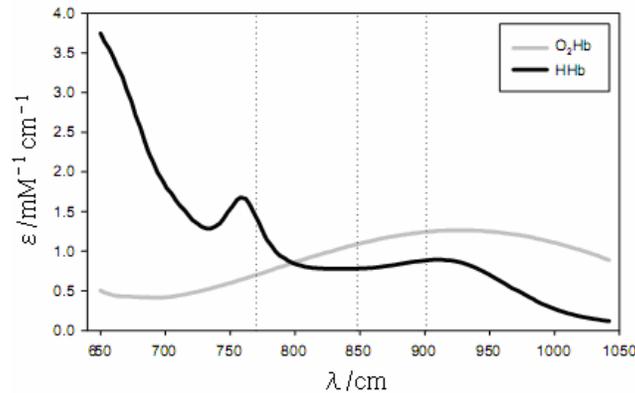
Figure 7. Extinction coefficient of the $O_2$Hb and HHb[10]

As shown in the figure, even though the concentration of HHb and O₂Hb in the muscle is about 100μM to the maximum, their absorption coefficient is in the range of 0.01～0.2cm$^{-1}$ at the wavelength range 700～1000nm of the light source. In this range the reduced scattering coefficient of the different muscles is in the range of 6～10cm$^{-1}$ and decreases gradually with the increasing of the wavelength[11].

We predicted the separation distance and DPF values suited to the analysis of the O₂Hb and HHb of the different concentrations in the muscle tissue from the attenuation data cube (Figure 2). (Table3).

Table3. Analysis of the separation distance and DPF values suited to the analysis of the O₂Hb and HHb ($\mu_s'$=6～10 cm$^{-1}$)

| [Hb]$_{max}$ | 100μM ($\mu_a$=0.01～0.2cm$^{-1}$) | | 50μM ($\mu_a$=0.01～0.1cm$^{-1}$) | | 25μM ($\mu_a$=0.01～0.05cm$^{-1}$) | |
|---|---|---|---|---|---|---|
| ρ/cm | $\partial^2 A/\partial \rho^2$ | DPF | $\partial^2 A/\partial \rho^2$ | DPF | $\partial^2 A/\partial \rho^2$ | DPF |
| 1～2 | -0.354 | 2.63 | -0.369 | 3.13 | -0.381 | 3.62 |
| 2～3 | -0.123 | 3.02 | -0.127 | 3.69 | -0.132 | 4.39 |
| 3～4 | -0.061 | 3,24 | -0.063 | 4.02 | -0.065 | 4.85 |
| 4～5 | -0.036 | 3.38 | -0.037 | 4.23 | -0.038 | 5.15 |

As shown in table3, the separation distance interval where the second derivative of the attenuation is situated near zero is between 3 and 5 and the values of DPF are placed between 3.5 and 5. Such results agree with the data of the separation distance and DPF used to determine the amount of the hemoglobin derivatives in the different kinds of muscle tissues[12]. (Table4).

Table 4. Separation distances and DPF values used for the analysis of the hemoglobin derivatives[12]

| muscle | method | analysis | Separation distance/cm | DPF |
|---|---|---|---|---|
| FDS | AO | O₂Hb | 4.0 | 3.5 |
| FDS | AO | Hb$_{dif}$ | 4.0 | 3.5 |
| BR | AO | Hb$_{dif}$ | 4.0 | 3.5 |
| FDS | VO | HHb | 4.0 | 5.0 |
| BR | VO | HHb | 4.0 | 3.5 |
| GASTR | AO | O₂Hb | 5.0 | 3.0 |
| BR | VO | Hb$_{dif}$ | 4.16 | 3.5 |
| BR | AO | Hb$_{dif}$ | 3.51-5.1 | 2.8-3.2 |
| BR | AO | Hb$_{dif}$ | 3.59 | 3.0-3.5 |
| SOL | AO | O₂Hb | 4.3 | 4.5 |

AO: arterial occlusion, VO: vena occlusion, Hb$_{dif}$= O₂Hb- HHb

## 4. Conclusion

We considered the problem of setting the source-detector separation distance arising when predicting the absorption coefficient from the reflection image of the tissue by the spatially resolved spectroscopy.

We calculated the attenuation data cube with the absorption coefficient, reduced scattering coefficient of tissue and separation distance of detector as its independent variables using the diffuse approximation resolution for the photon migration

We predicted the separation distance and DPF values suited to the spatially resolved spectroscopy by calculating and analyzing the attenuation slope with respect to the variables and the first derivative of the attenuation with respect to the absorption coefficient from the attenuation data cube, respectively.

When measuring the hemoglobin derivatives in the muscles, the proper separation distance interval is 3-5cm and DPF interval is 3.5-5 and these values agree well with the preceding experimental data.